\documentstyle[preprint,eqsecnum,aps]{revtex}

\begin{document}
\draft

\title{
	 	Dimensional Reduction and Quantum-to-Classical \\
   		Reduction at High Temperatures}

\author{	M.A. Stephanov%
\thanks{E-mail: misha@uiuc.edu.}\\
		Department of Physics, U. of Illinois
		at Urbana--Champaign,\\
		1110 W. Green St.,
		Urbana, IL 61801--3080, USA }

\date{}
\maketitle

\begin{abstract}

We discuss the relation between dimensional reduction in quantum
field theories at finite temperature and a familiar quantum
mechanical phenomenon that quantum effects become negligible at high
temperatures. Fermi and Bose fields are compared in this respect.
We show that decoupling of fermions from the dimensionally reduced
theory can be related to the non-existence of classical statistics
for a Fermi field.

\end{abstract}

\narrowtext

\section{Introduction}

The phenomenon of dimensional reduction in finite temperature field
theories has attracted considerable interest recently. This phenomenon
simplifies the description of finite temperature phase transitions
such as, for example, the chiral symmetry restoration transition in
QCD~\cite{PiWi84Wi93}, or the deconfinement transition in pure gauge
theories~\cite{SvYa82}. It helps to understand the behavior of QCD at
$T\gg T_c$ as well~\cite{ApPi81Na83La89Al89Re92}. The study of the
electroweak phase transition has been also focused on the issue of
dimensional reduction recently~\cite{Ka94}.

Let us first briefly present the common view on the dimensional
reduction existing in the literature.
The traditional description is based on
the Euclidean formalism. In this approach a given system
is living in a box of $d+1$ dimensions. The extent in $d$ spatial
dimensions is much larger than the physical scale (correlation length
in the system). The extent in the $d+1$-th (Euclidean time) dimension
is $1/T$, where $T$ is the temperature. The boundary conditions in
this time direction are periodic for Bose fields and antiperiodic
for Fermi fields.

Perturbation theory at finite temperature%
\footnote{
The argument of this paragraph applies to theories interacting
sufficiently weakly.
Also, masses of the particles are neglected compared to $T$.
}
contains integrals over spatial $d$--momenta and (infinite) sums over
discrete Matsubara frequencies $\omega_n$. Such a sum can be
interpreted as a sum over particles with masses $\omega_n$ in the
perturbation theory for some $d$-dimensional field theory at zero
temperature. One can argue then that the particles with high Matsubara
masses can be integrated out and low momentum physics can be described
by a $d$-dimensional effective theory for particles with smallest
Matsubara masses. The effect of other Matsubara modes is to
renormalize the couplings of this effective theory.
For a Bose field the Matsubara masses are quantized
as even multiples of $\pi T$.  Thus, one can describe physics at
momentum scales much less than $T$ by an effective $d$-dimensional
theory of the $\omega=0$ Matsubara mode.  Matsubara masses for
fermions are odd multiples of $\pi T$ and thus at the scale much less
than $T$ they decouple.

The dimensional reduction has a simple geometric origin. Fluctuations
with wavelength larger than the extent in the Euclidean time direction
$1/T$ are ``squeezed'' in this direction and behave like
$d$-dimensional fluctuations. In other words, for such fluctuations
the system looks like a ``pancake''. If such fluctuations determine
the physics of a phase transition (critical behavior) then this
physics is effectively $d$-dimensional.

In this paper we wish to emphasize the fact that these long wavelength
fluctuations are simply classical thermal fluctuations.
Such a view provides a nice physical interpretation of the dimensional
reduction as a well known phenomenon that quantum effects can be
neglected at high temperatures.  In essence, one uses the fact that
quantum oscillators with frequencies $\omega$ behave as classical
when%
\footnote{
Later on we shall use units in which $k_B=1$ and $\hbar=1$.
}
$\hbar\omega\ll k_B T$,
i.e., when thermal energies are much larger than the typical distance
between quantum levels.%
\footnote{
In quantum mechanics of one degree of freedom $x$ the reduction to
classical statistics at high temperatures can be viewed formally as a
dimensional reduction at $d=0$. Explicit relation between quantum and
classical partition functions at high $T$ is discussed in the path
integral formalism in~\cite{Fe72}. The functional integral over
trajectories $x(t)$ (imaginary $t$) in quantum partition
function reduces to an integral over $x$.
}
In quantum field theory the corresponding situation can occur either
as $T\to\infty$ or at a second order phase transition as $T\to T_c$
and $\omega\to 0$.

Such an interpretation of the dimensional reduction is missing in
the current literature on this
subject~%
\cite{PiWi84Wi93,SvYa82,ApPi81Na83La89Al89Re92,Ka94,RoSp94KoKo94}.
We do not attempt to rederive results of the traditional approach to
dimensional reduction.
The motivation for this note is to show a simple and rather familiar
physical picture behind the phenomenon which is largely viewed as a
technical trick.
In the context of condensed matter phenomena such a physical
picture is rather straightforward and quantum versus classical
transitions have been a subject of research~\cite{He76Yo94}. However,
in such a context the relation to the issue of dimensional reduction
is obscured by the absence of relativistic (space--time) invariance.

In Section \ref{sec:bose} we discuss how in the case of a free or
weakly interacting Bose field at finite temperature the classical
statistical behavior of low momentum modes is related to the
dimensional reduction.  Nothing is new there. We only wish to point
out the relation between the dimensional reduction and this well-known
statistical behavior of bosonic fields.  Similar discussion is given
for weakly interacting Fermi fields in Section \ref{sec:fermi}. We
relate the non-existence of classical statistics for Fermi fields to
the decoupling of fermions in the dimensionally reduced theory.

We discuss weakly coupled theories partly because the traditional
Euclidean Matsubara masses approach to dimensional reduction with
which we wish to make contact applies only then.  More general
discussion is given at the end of Section \ref{sec:con}.

\section{Bose fields}
\label{sec:bose}

Consider a quantum theory of some Bose field. For simplicity
consider a massless field. Such a limit is useful for studying
critical phenomena in scalar theories near the finite $T$ phase
transition, when the ($T$ dependent) mass is small compared to $T$. We
shall consider a noninteracting or weakly interacting theory.  In
fact, for our purposes a photon gas is a good example.

Our field theory can be viewed as a set of noninteracting (or very
weakly interacting) harmonic oscillators -- momentum modes -- with
frequencies $\omega$. At finite $T$ each oscillator has average
energy according to Planck:
\begin{equation}
\epsilon(\omega) = { \omega \over \exp(\omega/T) - 1 }.
\label{e}
\end{equation}
This corresponds to the average occupancy of each mode by
the photons (or whatever bosons):
\begin{equation}
n(\omega)= { 1  \over \exp(\omega/T) - 1 }.
\label{n}
\end{equation}

Now we note that oscillators with frequencies $\omega\ll T$ are
excited to very high energy levels $n\gg 1$. For such oscillators, as
we know, quantum effects are small. Thus the modes with $\omega\ll T$
can be described by a classical (statistical) theory. For example,
from equation~(\ref{e}) we can find that in such a limit
$\epsilon\approx T$ in accordance with a general theorem of classical
statistical mechanics (equipartition of energy).%
\footnote{A classical oscillator has two degrees of freedom:
the coordinate and the momentum.}
 For the photon gas this is the
Rayleigh--Jeans region of frequencies.  The classical
statistics/thermodynamics of a field represented by these modes is
what the dimensionally reduced theory is.

Where are the Matsubara frequencies in this picture? They are
given by the poles in the distribution (\ref{n}). They are imaginary
except $\omega=0$. This pole is responsible for the equipartition
of energy at small $\omega$: $\epsilon=\omega n(\omega)\approx T$.

The classical thermodynamics of the photon gas is
inconsistent because of the ultraviolet catastrophe: the total
energy (or, specific heat) diverges because of the contribution
of high frequency modes. This means that such a theory can be only
an {\em effective theory} for the modes with frequencies smaller than
$O(T)$. The UV divergence is regularized by quantum effects.

\section{Fermi fields}
\label{sec:fermi}

Next, consider a (free or weakly interacting) theory of fermionic
excitations. For reasons similar to the bosonic case we neglect the
mass of the fermions.  For a Fermi field the momentum modes cannot be
viewed as quantum oscillators, rather they are two-level systems. The
average energy and the occupation number of each mode are given by:
\begin{equation}
\epsilon_f(\omega) = { \omega \over \exp(\omega/T) + 1 },
\label{ef}
\end{equation}
\begin{equation}
n_f(\omega)= { 1  \over \exp(\omega/T) + 1 }.
\label{nf}
\end{equation}

The modes with small $\omega$ do not behave classically at
all. If such a mode could be represented by a classical system such
a system would have less than one degree of freedom:
$\epsilon\approx\omega/2$ for $\omega\ll T$. Of course, this is
due to the fact that a fermion level cannot be occupied by more
than one fermion: $n_f\le 1$, and coherent classical Fermi fields
do not exist. Again, the Matsubara frequencies are the
poles of $n_f(\omega)$. The absence of a pole at $\omega=0$ is the
reason for the non-classical behavior.

The critical behavior at a second order phase transition is determined
by long wavelength (low $\omega$) fluctuations. From (\ref{e}) and
(\ref{ef}) we see that the contribution of a Fermi field to the total
energy is much smaller than the contribution of a Bose one at very
small $\omega$.%
\footnote{
More precisely, a low frequency ($\omega_p\ll T$) bosonic mode
contributes to the free energy: $T\ln(\omega_p/T)$, while a fermionic
one contributes: $\omega_p/2-T\ln2$. The latter expression is the
energy minus the entropy for a two--level system. For the bosonic mode
the entropy dominates in the free energy.  If we take
$\omega_p^2=a(T-T_c)+p^2$ (Gaussian model) near a phase transition at
$T=T_c$, we find that the contribution to the specific heat
$C_V=-T(\partial^2F/\partial T^2)$ is proportional to $\omega_p^{-4}$
for bosons and to $\omega_p^{-3}$ for fermions.
}
This is related to the decoupling of fermions
in the formalism based on Matsubara masses.

What is the meaning of masslessness of these fermions then? One can
see explicitly that there is no pole at $p=0$ in the (free) fermion
propagator at finite $T$ for {\em massless} fermions. One can also
calculate the free propagator in the coordinate space and see that it
decays exponentially with spatial separation $r$ as $\exp(-T r/\pi)$.
What happens to the pole at $p=0$ which exists at zero $T$?

It is instructive to see this in the following way.  Consider a
(hermitian) perturbation $V$ creating/annihi\-lating a fermion with
momentum $\mbox{\boldmath $p$}$. The propagator is proportional to the
response of the system (e.g., change of the free energy) to this
perturbation.  The mode with momentum $\mbox{\boldmath $p$}$ is a
two-level system with energies: $E_0=0$ and $E_{1 }=|\mbox{\boldmath
$p$}|$. The levels are ``repulsed'' under the perturbation $V$. To the
order $V^2$ the shifts are given by:
\begin{equation}
\Delta E_0 =
{ \langle0|V|1 \rangle\langle1 |V|0\rangle \over  E_0 - E_1},
\qquad
\Delta E_{1 } =
{ \langle1 |V|0\rangle\langle0|V|1 \rangle \over E_1 - E_0 }
\label{e0 e1}
\end{equation}

The shift of $E_0$ is due to a virtual process in which the particle
is created from the vacuum and then annihilated. The shift of $E_{1 }$
is due to a process in which the particle is annihilated and then
created back. The amplitudes for these processes are equal because $V$
is hermitian. Hence the level shifts have equal magnitudes and
opposite signs.

Only the first process contributes to the propagator at zero
temperature.  The pole at $\mbox{\boldmath $p$}=0$ is due to the
one-massless-particle intermediate state $|1\rangle$.  At nonzero
temperature the second process (with vacuum intermediate state) also
contributes to the response of the system because there are particles
in the thermal bath already. Its contribution cancels the pole at
$\mbox{\boldmath $p$}=0$.  Indeed, the change in the partition
function $Z$ under the perturbation $V$ is given by%
\footnote{
The eqs.~(\ref{e0 e1}), (\ref{delta z}) constitute in essence the
spectral representation of the propagator. Using this representation
one can formally generalize the argument to theories where fermions
interact not necessarily weakly.
} :
\begin{equation}
-T\Delta Z =
\Delta E_0  e^{- E_0/T} + \Delta E_{1 } e^{- E_{1 }/T}=
- { |\langle0|V|1 \rangle|^2 \over |\mbox{\boldmath $p$}| }
(1-e^{- |\mbox{\boldmath $p$}|/T}).
\label{delta z}
\end{equation}

We see that although massless fermions are in the spectrum, they do
not produce a pole at $\mbox{\boldmath $p$}=0$ at finite $T$. Thus,
for example, massless fermions do not induce long range exchange
interactions at finite temperature.

\section{Conclusions and discussion}
\label{sec:con}

The purpose of this note is to point out that dimensional reduction in
quantum field theories can be understood as a familiar quantum
mechanical phenomenon: quantum effects become negligible at high
temperatures and classical statistics can be applied. The question of
whether there is dimensional reduction at a given phase transition at
finite $T$ or there is not (as it is discussed, for example, in the
case of the electroweak transition~\cite{Ka94}) is equivalent to the
following question: Is the physics of the transition classical or
quantum? In other words, can the transition be described by some
classical statistical field theory or not?

In pure bosonic theories one can expect the answer to be yes if the
transition is of the second order. The physics of such a transition is
determined by long wavelength (low frequency) fluctuations which are
classical. The corresponding classical statistical field theory is
what the dimensionally reduced theory will be.

In theories with fermions if the answer is to be yes,
then the fermions must decouple: fermionic excitations
are not classical.

If, for some reason, fermionic degrees of freedom are important at a
phase transition then the transition is not classical. Such a
possibility has been discussed recently for Yukawa and Gross-Neveu
models~\cite{RoSp94KoKo94}.  However, this possibility seems unlikely
at a {\it second} order phase transition at finite~$T$. The physics of
such a transition (critical behavior) is determined by low momentum
excitations.  Fermionic excitations of low momentum are suppressed by
Fermi statistics (as discussed in Section~\ref{sec:fermi}).

It might be possible to make a very general statement: Any second
order phase transition at finite temperature is classical. From this
point of view, the only possibility for transitions with
non-negligible quantum effects (and thus possible importance of
fermionic degrees of freedom) is provided by first order transitions
(or, obviously, by transitions at exactly zero~$T$).

Consider the hot electroweak phase transition as an example.  The
question of whether one can use dimensionally reduced theory to
describe the transition or one cannot is, from a physical point of
view, the question of whether the transition is classical or
not. Qualitatively, the answer is simple: if the transition is of the
second (or weakly first) order then the physics of the transition
is dominated by classical thermal fluctuations.  If the transition
is strongly first order (which it is if the Higgs mass at zero $T$ is
small) then quantum effects are important at $T_c$ and there is no
dimensional reduction.

The discussion in this paper was mostly limited to theories which are
free or interacting weakly.%
\footnote{
So that the representation of the field as a set of weakly
coupled oscillators (or two-level systems) is possible.
}
In this case the traditional Euclidean Matsubara masses argument shows
that a dimensionally reduced description exists for bosonic but not
for fermionic low momentum modes. This corresponds, as we saw,
to classical statistical behavior of bosonic low momentum modes
and non-classical behavior of fermionic ones.

In a strongly coupled theory it could, in principle, depend on the
given dynamics whether a dimensionally reduced Euclidean description
exists at high temperature (or at a phase transition). By such a
description we mean some {\em local} effective Euclidean theory in one
less dimension.  Simple arguments discussed here suggest that for
theories interacting sufficiently weakly (at the energy scale of the
order of $T$) such a description exists at a {\em second} order phase
transition at finite temperature. It remains to be seen whether this
conjecture holds for strongly coupled theories.

Nevertheless, it is clear that in {\em any} sensible case if a
dimensionally reduced Euclidean theory for low momentum modes does
exist, this reduced theory is simply a classical statistical
theory. It is an {\em effective} theory and a cutoff
of order $T$ should be implemented (like, e.g., in the UV
catastrophe). The effective classical degrees of freedom (collective
excitations), the thermal masses and the couplings, can non-trivially
depend on and should be derived from the underlying {\em quantum}
(thermo)dynamics.

In weakly coupled theories perturbation theory can be used for such a
derivation, although the resulting classical statistics (critical
behavior) is non-perturbative. A typical example is $\lambda\phi^4$
theory with $\lambda\ll1$. It is convenient to view the thermal mass
$m(T)$ as a measure of the distance from the criticality: $m(T)\to 0$
as $T \to T_c$. Low momentum modes become classical when $m(T)\ll T_c$
while perturbation theory still works as long as $m(T)\gg\lambda
T_c$. Similar situation occurs in the electroweak
theory~\cite{Ka94}. In strongly coupled theories, such as QCD at
$T_c$, the effective classical theory is not derivable in such a
perturbative way. Special methods and insights are necessary.
For example,
symmetry principles can be used to determine the relevant degrees of
freedom of the effective theory~\cite{PiWi84Wi93,SvYa82}.

\acknowledgments

The author thanks A. Kocic and J. Kogut for fruitful discussions and
M. Tsypin for valuable comments. The work was supported by the grant
NSF--PHY 92--00148.


\begin{references}


\bibitem{PiWi84Wi93}
		R. Pisarski and F. Wilczek, Phys. Rev. {\bf D29}, 338 (1984);
		F. Wilczek, Nucl. Phys. {\bf A565}, 123c (1994).

\bibitem{SvYa82}
		B. Svetitsky and L.G. Yaffe, Nucl. Phys.
		{\bf B210}, 423 (1982).

\bibitem{ApPi81Na83La89Al89Re92}
		T. Appelquist and R. Pisarski,
		Phys. Rev. {\bf D23}, 2305 (1981);
		S. Nadkarni, Phys. Rev. {\bf D27}, 917 (1983);
		N.P. Landsman, Nucl. Phys. {\bf B322}, 498 (1989);
		R.F. Alvarez-Estrada, Physica {\bf A158}, 178 (1989);
		T. Reisz, Z. Phys. {\bf C53}, 169 (1992).

\bibitem{Ka94}
		For a review of electroweak phase transition at finite
		temperature and references see, e.g.,
		K. Kajantie, Hot Electroweak Matter,
		a plenary talk at the `Lattice 94' Conference,
		Bielefeld Sept. 27 -- Oct. 1, 1994, to appear in Nucl.
		Phys. B (Proc. Suppl.), hep-lat/9412072.

\bibitem{Fe72}
		R.P. Feynman, Statistical Mechanics (W.A. Benjamin, Inc.,
		Reading, Massachusetts, 1972) Chapter 3.

\bibitem{RoSp94KoKo94}
		B. Rosenstein, A.D. Speliotopoulos and H.L. Yu,
		Phys. Rev. {\bf D49}, 6822 (1994);
		A.~Kocic and J. Kogut,
		Phys. Rev. Lett. {\bf 74}, 3109 (1995).

\bibitem{He76Yo94}
		See, e.g., J.A. Hertz, Phys. Rev. {\bf B14}, 1165 (1976); for
		a review see, e.g.,
		A.P. Young, Quantum Phase Transitions,
		a plenary talk at the `Lattice 94' Conference,
		Bielefeld Sept. 27 -- Oct. 1, 1994, to appear in Nucl.
		Phys. B (Proc. Suppl.).


\end{references}
\end{document}